# An Implementation Framework (IF) for the National Information Assurance and Cyber Security Strategy (NIACSS) of Jordan

Ahmed Otoom[1] and Issa Atoum[2]
[1]National Information Technology Center, Jordan
[2]Information Technology Department, Philadelphia University, Jordan

**Abstract:** *This paper proposes an implementation framework that lays out the ground for a coherent, systematic, and comprehensive approach to implement the National Information Assurance and Cyber Security Strategy (NIACSS) of Jordan. The Framework 1). Suggests a methodology to analyze the NIACSS, 2). Illustrates how the NIACSS analysis can be utilized to design strategic moves and develop an appropriate functional structure, and 3). proposes a set of adaptable strategic controls that govern the NIACSS implementation and allow achieving excellence, innovation, efficiency, and quality. The framework, if adopted, is expected to harvest several advantages within the following areas: information security implementation management, control and guidance, efforts consolidation, resource utilization, productive collaboration, and completeness. The framework is flexible and expandable; therefore, it can be generalized.*

## 1. Introduction

The National Information Assurance and Cyber Security Strategy (NIACSS) developed recently by the Ministry of Information and Communications Technology (MoICT) of Jordan represents only the first step. The Government of Jordan (GoJ) will most likely be confronted with challenges that will emerge from the implementation of this Strategy. According to Alrawabdeh [1], the Arab World is facing challenges related to a number of social, technological, financial and legal issues. As a result, there is an inevitable need to develop well-defined plans that anticipate and address these challenges before they are triggered.

According to the MoICT, the NIACSS is intended to augment the overall national security strategy for Jordan. The NIACSS identifies strategic objectives and national priorities. The strategic objectives aim to: strengthen National security, minimize risks to Critical National Infrastructure (CNI), minimize damage and recovery time, enhance economy and National prosperity, and increase Cyber Security and Information Assurance (CS&IA) awareness. National priorities address the critical needs required to guide the implementation towards achieving the National objectives. The national priorities cover the following areas: risk management, Jordan Computer Emergency Response Team (JO-CERT), awareness, standards and policies, international cooperation, securing national information systems and networks, CNI protection, National Encryption Center (NEC), and legal regulatory regime. A successful implementation will demand collaboration within government, with international partners, with the private sector, and with the citizenry of Jordan [12].

Most of the international information security strategies include guidelines in order to facilitate their implementation [5, 7, 16]. Phahlamohlaka *et al.* [14] suggest an Awareness Toolkit as an approach to implement the strategy of South Africa. Other researchers tackle strategy implementation from different perspectives and to different extents, for example proper controlling is suggested to ensure security governance [15], and Information Security Management System (ISMS) standards [8] and ISMS evaluation [6] are being continually explored to their crucial importance to CS&IA. Most of the international efforts are dedicated to their countries or to specific organizations within these countries; hence, without major alteration, most of these efforts cannot be generalized or be globally reused.

This paper considers the NIACSS to only represent the first step towards achieving the expected objectives. McConkey [11] suggests that even the most technically perfect strategic plan will serve little purpose if it is not implemented. According to a survey published by (KPMG), a large professional service network, 69% of project failures are due to improper implementation of project management methodologies [17]. The implementation of the NIACSS is no exception; thus top level management must commit to a persistent implementation mechanism utilizing the appropriate resources in order to fulfil the NIACSS implementation. The implementation should be managed utilizing a proper methodology; hence this paper tries to add value in this area.

This paper tries to answer the major question that will be triggered just after the anticipated approval of the NIACSS; "What is the implementation framework that the GoJ is going to adopt in order to implement the NIACSS?" Therefore, this paper presents a proposed Implementation Framework (IF) that lays out the ground for a coherent, systematic, and comprehensive approach to implement the NIACSS. The IF consolidates the implementation efforts, suggests a set of strategic controls that enable decision makers to have full control on the implementation process that empowers them to take the right decisions once these decisions are needed, facilitates the implementation of the NIACSS and engages the involved parties into a productive collaboration as quickly as possible; otherwise the NIACSS will become obsolete as time goes on. By adopting the IF, involved parties will follow a well-defined systematic approach towards achieving the strategic objectives. Moreover, the IF makes the implementation of the NIACSS manageable by breaking it down into smaller components utilizing the "divide and conquer" principle. The IF is flexible and expandable so it can be generalized rather than being limited to NIACSS.

First, the IF is introduced and broken down into several major components that are illustrated with examples where applicable. Then, the paper is concluded and a list of possible topics for future research is provided, respectively.

## 2. Implementation Framework

According to Fred David, A strategic planning process shown in Figure 1 consists of the following: strategy formulation, strategy implementation, and strategy evaluation [3].

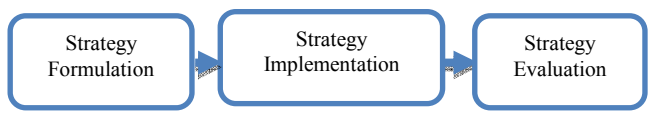

Figure 1. Strategic planning process [3].

Successful strategy formulation does not necessarily result in successful strategy implementation [4]. The IF fits in the strategy implementation phase; strategy formulation and evaluation are outside the scope of this paper. The high level block diagram of the IF is shown in Figure 2. It serves to give a broad view of the IF. The IF consists of four major components:

- NIACSS Analysis.
- National Information Assurance and Cyber Security Agency (NIACSA).
- Strategic Moves.
- Objectives.

The IF essentially helps transforming the National CS&IA from the current state to the future state. Both current and future states related to CS&IA are directly or indirectly documented in the NIACSS; though further analysis is required to make knowledge about these states more valuable and understandable. Initially, the IF proposes a methodology to analyze the NIACSS and break it down into well-defined components. Then, the analysis results are used to:

1. Guide the design and the establishment of an entity responsible for the strategy implementation; called the NIACSA.
2. Determine the strategic moves that are necessary to achieve the national objectives.

Finally, the IF suggests that the NIACSA, develop and deploy a set of controls to govern the transformation process. The NIACSS has already anticipated a critical need to establish the NIACSA to act as a focal entity responsible for the implementation of the NIACSS, but the details of this entity are left to the implementation phase. Later on, in section 2.2 of this paper, we will see how the IF facilitates the establishment of this entity. The following subsections will explore the major components of the IF in more details.

### 2.1. NIACSS Analysis

The rationale behind NIACSS analysis is as follows:

1. We need to break the NIACSS into manageable understandable components; usually those who implement the strategy likely be different people from those who formulate it. Therefore, care must be taken so that the NIACSS implementers understand the strategy and the reasoning behind it. Otherwise, the implementation might not succeed or faced with resistance.
2. The output of the NIACSS analysis will greatly determine the organizational structure of the NIACSA.
3. We need to be able to determine the strategic moves that map the NIACSS towards achieving the identified objectives; so we guarantee a full coverage of the NIACSS and eliminate any possible redundancy across these strategic moves.

To accomplish the analysis, we apply the concept of "Viewpoints" that is being used in software engineering to gather and validate requirements; for more details refer to [9, 10, 13].

As shown in Figure 2, The NIACSS is taken as an input to the analysis process. The analysis team may include, but is not limited to, members from: MoICT, National Information Technology Center (NITC),

Internet Service Providers (ISPs), health sector, IT business experts, and security departments. It may also be useful to include team members from people who participated in the NIACSS development. The more professional and diverse the team, the more successful the analysis output will be. The "viewpoints" of the team are gathered, incorporated, and summarized. The analysis team must resolve conflict, generate a reconciled understanding, and make sure that analysis is complete. An example on a "viewpoints" technique applied to the NIACSS is shown in Figure 3, it is given to illustrate the point and it is not meant to be thorough nor comprehensive.

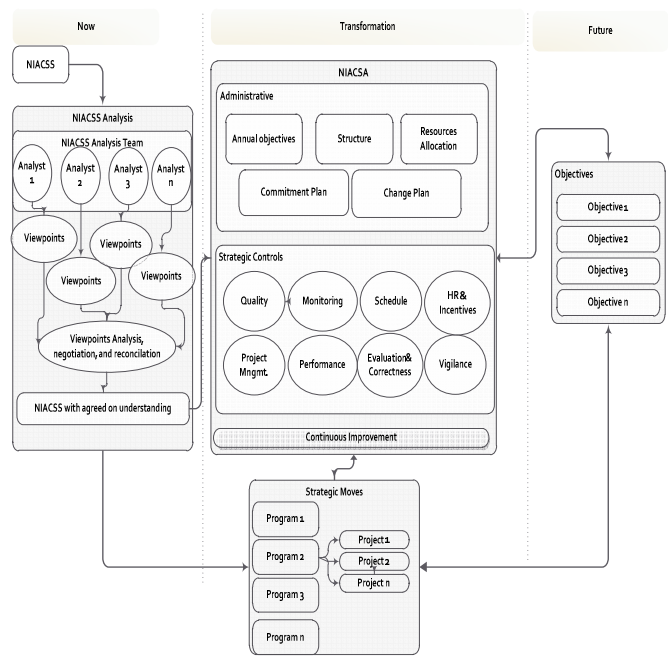

Figure 2. Proposed implementation framework.

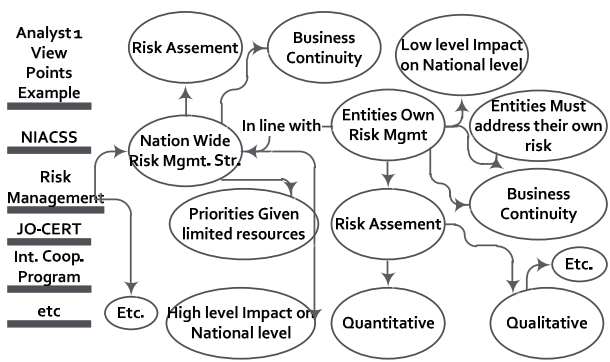

Figure 3. Example on "Viewpoints" applied to NIACSS.

## 2.2. NIACSA

The NIACSA will be the major entity responsible for the implementation of the NIACSS. This entity does not exist yet and it has to be established. The NIACSA is foreseen as a central national entity for governmental and non-governmental organizations regarding all information assurance and cyber security related issues. The transition from strategy formulation to strategy implementation shifts the responsibility to this entity. This arises several central implementation needs that include, but not limited to: annual objectives, organizational structure, allocating resources, commitment plan, change plan, strategic controls, and continuous improvement. We will discuss all of these needs in the following subsections except for the strategic controls and continuous improvement that are discussed later on in sections 2.4 and 2.5, respectively.

### 2.2.1. Annual Objectives

The objectives listed in the NIACSS are long term objectives that may take years to achieve. Therefore, we suggest breaking these long term objectives down into annual objectives. Annual objectives will help the NIACSA allocate resources gradually and expand its organizational structure as needed. Moreover, the NIACSA will be able to assist its performance and measure its progress towards achieving long term objectives via incrementally achieving these annual objectives.

### 2.2.2. NIACSA Organizational Structure

Bernard [2] proposes that organization structure should follow strategy and should be crafted to facilitate the realization of the strategy. David [4] illustrates that changes in strategy lead to changes in organizational structure and structure should be designed to facilitate the strategic pursuit of a firm and, therefore, follow strategy. This paper suggests that structure should empower the employees and engage them into a collaborative environment. It must be adaptable to help the NIACSA be effective and able to provide CS&IA products and services in an efficient manner.

The NIACSS has already anticipated the need to establish the NIACSA, but no implementation details about this entity were given. The NIACSS suggests changes in the way that the nation should approach CS&IA. Therefore, the organization structure should enable the NIACSA to allocate the required resources in order to achieve the NIACSS objectives.

This paper highlights how the organization structure is critically related to the IF. There are 7 basic types of organizational structures: functional, divisional by geographic area, divisional by product, divisional by customer, divisional process, Strategic Business Unit (SBU), and matrix. We will only take the functional structure as an example to illustrate the point. The functional structure is a centralized structure that is most widely used because it is the simplest and least expensive of the seven alternatives. Building a detailed organization structure that considers the 7 basic types is outside the scope of this paper and is left for future research.

Figure 4 shows the organizational structure as a function of strategy; not the reverse. The mapping process should guarantee that the identified outputs of the analysis phase including: functions, assessments, objectives, and strategic moves are mapped into the organizational structure. While organizational structure must enable strategy, it must also take into account the pragmatic issues of culture, management style, reward systems, administrative and strategic controls, and continuous improvement. It might also be useful to look into some similar international organizations to reuse, customize or build upon existing experience.

Figure 5 shows an example of a possible high level organizational structure for the NIACSA.

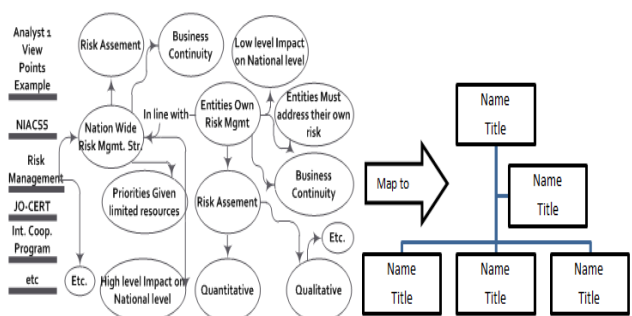

Figure 4. Organization structure as a function of strategy.

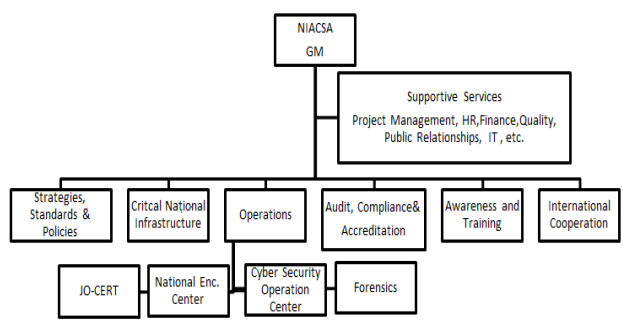

Figure 5. Example-high level organizational structure for NIACSA.

### 2.2.3. Resource Allocation

The NIACSA will need to allocate four types of resources: human resources, financial resources, technological and physical resources. Without effective resource allocation, the strategy execution will render to failure. Resource allocation should not be based on political or personal factors. The NIACSA should be able to allocate resources to fill the positions created by the new structure. Moreover, resource allocation can be gradually performed according to the annual objectives 2.2.1. Although, effective resource allocation is a major factor, it will not guarantee strategy implementation success that depends on many other factors like controls and commitment that efficiently utilize resources towards achieving the required objectives. Resource allocation for the NIACSA is suggested to be covered in dedicated studies; hence it is suggested as one of future research topics.

### 2.2.4. Commitment Plan

Successful strategy implementation is firmly linked to commitment. The human part in the implementation process makes a difference. Commitment plan will demonstrate management's commitment to strategy implementation. For the implementation to succeed, human issues of building and sustaining commitment must be addressed to minimize resistance. Individuals should not be left skeptical or jaded. The NIACSA must first build the commitment of leadership teams that will influence and shape the commitments of the staff. Individuals and NIACSA departments should be willing to sacrifice and work enthusiastically to meet the challenges that will inevitably emerge just after the establishment of the NIACSA.

### 2.2.5. Change Plan

The NIACSS implementation will have to bring different entities to cooperate and engage in a consolidated effort towards achieving the required objectives. These entities include, but not limited to: governmental organizations, private sector, and probably some international entities. Each of these entities will have its own goals and may try to resist changes that are not of their interests. Therefore, a change plan will anticipate and minimize this possible resistance to change. Failing to have such a plan may allow likely overt and covert resistance result in strategy implementation failure.

### 2.3. Strategic Moves

Strategic moves are actions taken to achieve one or more objectives. Strategic Moves are prescriptive and purposeful; they identify exactly what is to be done and directly act to achieve their objectives. Strategic moves must not contradict each other; rather, they should complement.

This paper suggests that Strategic Moves are designed as a set of coherent implementation programs. Each implementation program is broken into a set of complementary manageable tactical projects. After identifying the set of programs and projects, they are prioritized and ordered. When projects are independent, we suggest using a tool such as the matrix shown in Figure 6. However, for interdependent projects, we suggest using PERT charts such as the one shown in Figure 7.

|  | High | 4 | 5 | 6 |
|---|---|---|---|---|
| Cost | Medium | 2 | 3 | 5 |
|  | Low | 1 | 2 | 2 |
|  |  | High | Medium | Low |
|  |  | Priority (Pay Off) | | |

Figure 6. Independent project ordering.

Figure 8 shows an Example on a partial set of strategic moves. This example is given to clarify how the concept of strategic moves is applied to the NIACSS; it is not intended to be thorough nor comprehensive. This Example selects the awareness and training program from a set of available programs and breaks it down into a set of projects such that each project contributes to the achievement of one or more strategic objectives and each objective is achieved by implementing one or more projects.

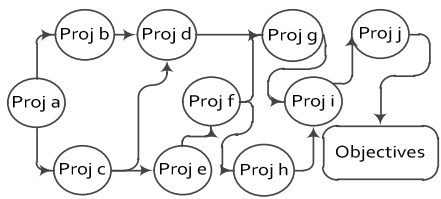

Figure 7. Interdependent projects ordering.

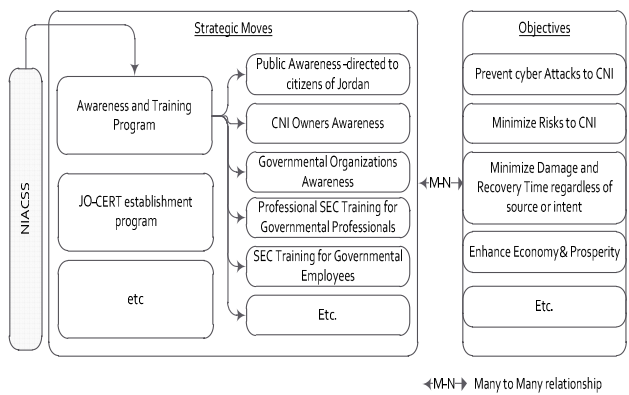

Figure 8. Strategic moves example.

### 2.4. Strategic Controls

The NIACSA should deploy a set of applicable strategic controls that are considered, from IF's point of view, very significant to the success of the NIACSS implementation. Strategic controls should allow decision makers determine whether the NIACSA is achieving innovation, efficiency, and quality. Moreover, they will enable decision makers make any necessary adjustments and improvements as early as possible in the implementation process. These Controls should be adaptable to the culture and they should evolve with the NIACSA. The NIACSA should be able to add, enhance, or delete controls as needed. The set of strategic controls may include, but not limited to: quality, monitoring, schedule, human resources and incentives, project management, performance, evaluation and correctness, vigilance, etc.

Quality controls will be applied throughout the whole process to make sure that plans are complete, correct, and best possible for the NIACSA. The NIACSA will need to monitor the implementation and peruse periodic reviews to determine its performance and assess the progress. The implementation of the NIACSS should be time bounded which triggers the need to deploy scheduling controls. The NIACSA will have to develop and utilize incentive measures to encourage employees to work towards achieving the identified objectives. Vigilance will enable the NIACSA to proactively scan the environment in order to deal with unanticipated events of strategic value.

These new or unforeseen events may make it necessary to change ongoing plans. As the environment and requirements change overtime, corrective actions will be necessary to guarantee that the implementation process is effective in reaching what the NIACSS is set out to achieve. The bottom line is that the NIACSA should deploy all necessary strategic controls that enable it to efficiently manage and control the implementation of the NIACSS towards achieving the required objectives within specified timeframes.

### 2.5. Continuous Improvement

The NIACSA should be committed to an ongoing continuous improvement process to improve all of its possible products, services, or processes.

### 2.6. Objectives

The NIACSA must deploy all necessary administrative and strategic controls to achieve the national objectives identified in the NIACSS. These objectives will guide the planning, expose priorities, and form a basis for organization, collaboration and evaluation. For the sake of completeness, here we want to highlight two major issues: long term objectives should be broken down into annual objectives (section 2.2.1) and there is a need to come up with more accurate measures to these objectives so that implementation progress can be assessed and controlled.

## 3. Future Research

Implementing the NIACSS is a big challenge; both technically and financially. The GoJ needs to support researchers in this regard for an efficient implementation within limited resources. This paper pinpoints several areas that may require further research:

- Perform a complete analysis for the NIACSS according to the framework described in section 2.1.
- Build a complete and detailed organizational structure for the NIACSA by considering:
  a. Different types of organizational structures.
  b. Duties and jobs description.
  c. Scope and size.
  d. Resource allocation.
  e. Working processes and procedures.
- Develop Change and Commitment Plans.
- Investigate possible outsourcing to several components of the NIACSA.
- Build a financial plan.

## 4. Conclusions

This paper presents an IF that aims to guide the implementation efforts of the NIACSS. It proposes a coherent systematic IF that tackles the implementation of the NIACSS as a whole.

The framework starts by analyzing the NIACSS. Then, it shows how the output of the analysis phase is used to guide the establishment of the NIACSA and the design of strategic moves necessary to achieve the national objectives. It also suggests and discusses several strategic controls necessary to govern the execution of the strategic moves towards achieving the required objectives. The set of strategic controls includes, but not limited to: quality, monitoring, schedule, incentives, project management, performance, evaluation, and vigilance.

The framework, if adopted, is expected to harvest several advantages within these areas: implementation management, control and guidance, efforts consolidation, resource utilization, productive collaboration, and completeness.

Although, this framework is designed to implement the NIACSS, we believe it is flexible, expandable, and can be generalized. Other strategies and corresponding strategic moves can be added easily without vital changes to the baseline framework.

## Disclaimer

This paper does not represent the thoughts, intentions, plans or strategies of the NITC, Jordan's MoICT, or any other Governmental or nongovernmental entity; it is solely the opinion of the authors. The NITC, MoICT, and/or any other entities are not responsible for the accuracy of any of the information supplied herein.

## Acknowledgements

The authors would like to thank the NITC for giving the first author of this paper a key role in the development and the implementation of the NIACSS. We are also grateful to Eng. Fahd Batayneh and Dr. Mohammad Al-Hammouri for their valuable comments and fruitful discussions. Last but not least, we wish to thank Miss Auhood Majali for editing this paper.

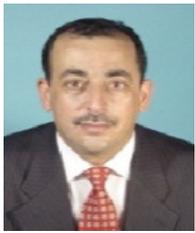

**Ahmed Otoom** is the DG's Advisor for the Implementation of the E-Gov Information Strategies and Policies, National Information Technology Center, Jordan. He received his PhD degree in computer science from Amman Arab University in 2007, dual MS degrees in computer science and information technology management from the Naval Postgraduate School, USA in 2000 and a BS degree in Computer Science from Mutah University in 1992. He has more than 19 years of experience in project management and IT-related projects. During 1992-2010, he worked as a system analyst, developer, and researcher for the IT Directorate in Royal Jordanian Air Force. During the same period, he also taught many computer science classes in the Prince Feisal Technical College and Alzaytoonah University of Jordan, respectively. His major areas of interest include computer security, operating systems and interoperability in large heterogeneous information system

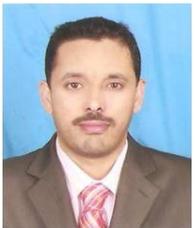

**Issa Atoum** is a Master Student at Philadelphia University, Jordan. He has more than 15 years of professional experience in IT services, project management and quality assurance. He is a certified Project Manager Professional (PMP®), ITIL®V3 and ISO/IEC 20000. He worked as a project manager for the government of Abu Dhabi and Dubai, UAE for mid and large scale projects in the domain of security and IT services. His major areas of interest are: computer security, semantic web and e-government.